\RequirePackage{ifpdf}
\ifpdf 
\documentclass[pdftex]{sigma}
\else
\documentclass{sigma}
\fi

\begin{document}

\allowdisplaybreaks

\renewcommand{\PaperNumber}{072}

\FirstPageHeading

\ShortArticleName{A Jacobson Radical Decomposition of the Fano-Snowf\/lake
Conf\/iguration}

\ArticleName{A Jacobson Radical Decomposition\\ of the  Fano-Snowf\/lake
Conf\/iguration}

\Author{Metod SANIGA~$^\dag$ and Petr PRACNA~$^\ddag$}

\AuthorNameForHeading{M. Saniga and P. Pracna}

\Address{$^\dag$~Astronomical Institute, Slovak Academy of
Sciences,\\
\hphantom{$^\dag$}~SK-05960 Tatransk\' a Lomnica, Slovak Republic}
\EmailD{\href{mailto:msaniga@astro.sk}{msaniga@astro.sk}}

\Address{$^\ddag$~J. Heyrovsk\' y Institute of Physical Chemistry,
v.v.i., Academy of Sciences of the Czech Republic,\\
\hphantom{$^\ddag$}~Dolej\v skova
3, CZ-18223 Prague 8, Czech Republic}
\EmailD{\href{mailto:pracna@jh-inst.cas.cz}{pracna@jh-inst.cas.cz}}

\ArticleDates{Received July 14, 2008, in f\/inal form October
17, 2008; Published online October 24, 2008}

\Abstract{The Fano-Snowf\/lake, a specif\/ic conf\/iguration associated with the smallest ring of ternions $R_{\diamondsuit}$ (\href{http://arxiv.org/abs/0803.4436}{arXiv:0803.4436} and \href{http://arxiv.org/abs/0806.3153}{arXiv:0806.3153}), admits an interesting partitioning with respect to the Jacobson radical of $R_{\diamondsuit}$. The totality of 21 free cyclic submodules generated by non-unimodular
vectors of the free left $R_{\diamondsuit}$-module $R_{\diamondsuit}^{3}$ is shown to split into three disjoint sets of cardinalities 9, 9 and 3 according as the number of Jacobson radical
entries in the generating vector is 2, 1 or 0, respectively. The corresponding ``ternion-induced'' factorization of the lines of the Fano plane sitting in the middle of the Fano-Snowf\/lake
is found to {\it differ fundamentally} from the natural one, i.e., from that with respect to the Jacobson radical of the Galois f\/ield of two elements.}

\Keywords{non-unimodular geometry over rings; smallest ring of ternions; Fano plane}

\Classification{51C05; 51Exx}

Projective lattice geometries over unital
associative rings $R$ (see, e.g., \cite{bgs} and references
therein) represent a very important generalization of classical
(f\/ield) projective spaces, being endowed with a number of
remarkable features not exhibited by the latter.  One of the most
striking dif\/ferences is, for certain $R$, the existence of {\it
free} cyclic submodules generated by {\it non}-unimodular vectors
of the free left $R$-module $R^{n+1}$, $n \geq 1$. In a couple of
recent papers \cite{shpp,hs}, an in-depth
analysis has been performed of such non-unimodular portions of the lattice geometries
when $R$ is the ring of ternions, i.e., a ring isomorphic to
that of upper triangular $2\times 2$ matrices with entries from an
arbitrary commutative f\/ield $F$. It has been found that for any $n
\geq 2$ these non-unimodular free cyclic submodules of $R^{n}$ can
be associated with the lines of $PG(n,F)$, the $n$-dimensional
projective space over $F$ sitting in the middle of such a
non-unimodular world. In the f\/inite case, $F = {GF}(q)$, basic combinatorial properties of
such conf\/igurations have been derived and
illustrated in exhaustive detail for the simplest, $n=q=2$ case~-- dubbed the
Fano-Snowf\/lake geometry. In the present paper we shall have
another look at the Fano-Snowf\/lake and show that this geometry admits an
intriguing decomposition with respect to the Jacobson radical of
the ring in question.

\begin{table}[t]
\centering
\caption{Addition ({\it left}) and multiplication ({\it right}) in
$R_{\diamondsuit}$.} \vspace*{0.2cm}
\begin{tabular}{||c|cccccccc||}
\hline \hline
$+$ & 0 & 1 & 2 & 3 & 4 & 5 & 6 & 7 \\
\hline
0 & 0 & 1 & 2 & 3 & 4 & 5 & 6 & 7 \\
1 & 1 & 0 & 6 & 7 & 5 & 4 & 2 & 3 \\
2 & 2 & 6 & 0 & 4 & 3 & 7 & 1 & 5 \\
3 & 3 & 7 & 4 & 0 & 2 & 6 & 5 & 1 \\
4 & 4 & 5 & 3 & 2 & 0 & 1 & 7 & 6 \\
5 & 5 & 4 & 7 & 6 & 1 & 0 & 3 & 2 \\
6 & 6 & 2 & 1 & 5 & 7 & 3 & 0 & 4 \\
7 & 7 & 3 & 5 & 1 & 6 & 2 & 4 & 0 \\
\hline \hline
\end{tabular}~~~~~
\begin{tabular}{||c|cccccccc||}
\hline \hline
$\times$ & 0 & 1 & 2 & 3 & 4 & 5 & 6 & 7  \\
\hline
0 &  0 & 0 & 0 & 0 & 0 & 0 & 0 & 0 \\
1 &  0 & 1 & 2 & 3 & 4 & 5 & 6 & 7 \\
2 &  0 & 2 & 1 & 3 & 7 & 5 & 6 & 4 \\
3 &  0 & 3 & 5 & 3 & 6 & 5 & 6 & 0 \\
4 &  0 & 4 & 4 & 0 & 4 & 0 & 0 & 4 \\
5 &  0 & 5 & 3 & 3 & 0 & 5 & 6 & 6 \\
6 &  0 & 6 & 6 & 0 & 6 & 0 & 0 & 6 \\
7 &  0 & 7 & 7 & 0 & 7 & 0 & 0 & 7 \\
\hline \hline
\end{tabular}\label{table1}
\end{table}

To this end, we f\/irst collect the necessary background information from \cite{shpp,hs}.
We consider an associative ring with unity $1$ $(\neq 0)$, $R$, and denote the free
left $R$-module on $n+1$ generators over $R$  by $R^{n+1}$. The set
$R(r_1, r_2,\dots, r_{n+1})$, def\/ined as follows
\begin{equation*}
R(r_1, r_2,\dots, r_{n+1}):= \left\{ (\alpha r_1, \alpha r_2,\dots, \alpha  r_{n+1}) \;|\;  \alpha \in R \right\},
\end{equation*}
is a {\it left} cyclic submodule of $ R^{n+1}$. Any such submodule
is called {\it free} if the mapping $\alpha \mapsto (\alpha r_1,
\alpha r_2,\dots, \alpha  r_{n+1}) $ is injective, i.e., if
$(\alpha r_1, \alpha r_2,\dots, \alpha  r_{n+1}) $ are all {\it
distinct}. Next, we shall call a vector $(r_1, r_2,\dots, r_{n+1}) \in
R^{n+1}$ {\it uni}modular if there exist elements $x_1,x_2,\dots, x_{n+1}$ in $R$ such that
\begin{equation*}
r_1 x_1 + r_2 x_2 + \cdots + r_{n+1} x_{n+1} = 1.
\end{equation*}
It is a very well-known fact (see, e.g., \cite{v81,v95,her,bh}) that if $(r_1,
r_2,\dots, r_{n+1})$ is unimodular, then $R(r_1, r_2,\dots,
r_{n+1})$ is free; any such free cyclic submodule represents a
point of the $n$-dimensional projective space def\/ined over~$R$~\cite{v95}. The converse statement, however, is not generally true. That
is, there exist rings which also give rise to free cyclic
submodules featuring exclusively non-unimodular vectors.  The f\/irst case where
this occurs is the smallest (non-commutative) ring of ternions,
$R_{\diamondsuit}$:
\begin{equation*}
   R_{\diamondsuit}  \equiv \left\{ \left(
\begin{array}{cc}
a & b \\
0 & c \\
\end{array}
\right)  \mid   a, b, c \in GF(2) \right\},
\end{equation*}
where the addition and multiplication is that of matrices over $GF(2)$.
From this def\/inition it is readily seen that the ring contains two maximal (two-sided) ideals,
\begin{equation*}
   I_1 =   \left\{ \left(
\begin{array}{cc}
0 & b \\
0 & c \\
\end{array}
\right) \mid  \,  b, c \in GF(2) \right\}
\qquad \mbox{and}\qquad
   I_2 = \left\{ \left(
\begin{array}{cc}
a & b \\
0 & 0 \\
\end{array}
\right) \mid  \, a, b \in GF(2) \right\},
\end{equation*}
which give rise to a non-trivial (two-sided) Jacobson radical $J$,
\begin{equation*}
   J = I_1 \cap I_2 = \left\{ \left(
\begin{array}{cc}
0 & b \\
0 & 0 \\
\end{array}
\right) \mid  \, b \in GF(2) \right\}.
\end{equation*}
Since for our further purposes it will be more convenient to work
with numbers than matrices, we shall relabel the elements of
$R_{\diamondsuit}$ as follows
\begin{gather*}
 0 \equiv \left(
\begin{array}{cc}
0 & 0 \\
0 & 0
\end{array}
\right),\qquad 1 \equiv \left(
\begin{array}{cc}
1 & 0 \\
0 & 1
\end{array}
\right),\qquad 2 \equiv \left(
\begin{array}{cc}
1 & 1 \\
0 & 1
\end{array}
\right),\qquad
3 \equiv \left(
\begin{array}{cc}
1 & 1 \\
0 & 0
\end{array}
\right), \nonumber \\
 4 \equiv \left(
\begin{array}{cc}
0 & 0 \\
0 & 1 \\
\end{array}
\right),\qquad 5 \equiv \left(
\begin{array}{cc}
1 & 0 \\
0 & 0
\end{array}
\right),\qquad  6 \equiv \left(
\begin{array}{cc}
0 & 1 \\
0 & 0
\end{array}
\right),\qquad
7 \equiv \left(
\begin{array}{cc}
0 & 1 \\
0 & 1
\end{array}
\right).
\end{gather*}
In this compact notation the addition and multiplication in the ring reads as shown in
Table~\ref{table1}. The two maximal ideals now acquire the form
\begin{equation*}
I_1 : = \left\{0, 4, 6, 7 \right\}
\qquad \mbox{and}\qquad I_2 : = \left\{0, 3, 5, 6 \right\},
\end{equation*}
and the Jacobson radical reads,
\begin{equation*}
J = I_1 \cap I_2 = \left\{0, 6 \right\}.
\end{equation*}

There exist altogether 21 free cyclic submodules of $R_{\diamondsuit}^{3}$ which are generated by non-unimodular vectors. Taking their complete list from  \cite{shpp} one sees that they
can be separated into the following three disjoint sets
\begin{gather*}
R_{\diamondsuit}(6,6,7)=R_{\diamondsuit}(6,6,4) \\[0.2ex]
\phantom{R_{\diamondsuit}(6,6,7)}{} = \left\{(0,0,0), (6,6,7), (6,6,4), (6,6,0), (0,0,4), (6,6,6), (0,0,6), (0,0,7) \right\},\\[0.2ex]
R_{\diamondsuit}(6,7,6)=R_{\diamondsuit}(6,4,6) \\[0.2ex]
\phantom{R_{\diamondsuit}(6,7,6)}= \left\{(0,0,0), (6,7,6), (6,4,6), (6,0,6), (0,4,0), (6,6,6), (0,6,0), (0,7,0) \right\},\\[0.2ex]
R_{\diamondsuit}(7,6,6)=R_{\diamondsuit}(4,6,6)\\[0.2ex]
\phantom{R_{\diamondsuit}(7,6,6)} = \left\{(0,0,0), (7,6,6), (4,6,6), (0,6,6), (4,0,0), (6,6,6), (6,0,0), (7,0,0) \right\},\\[0.2ex]
R_{\diamondsuit}(0,6,7)=R_{\diamondsuit}(0,6,4)\\[0.2ex]
\phantom{R_{\diamondsuit}(0,6,7)} = \left\{(0,0,0), (0,6,7), (0,6,4), (0,6,0), (0,0,4), (0,6,6), (0,0,6), (0,0,7) \right\},\\[0.2ex]
R_{\diamondsuit}(0,7,6)=R_{\diamondsuit}(0,4,6)\\[0.2ex]
\phantom{R_{\diamondsuit}(0,7,6)} = \left\{(0,0,0), (0,7,6), (0,4,6), (0,0,6), (0,4,0), (0,6,6), (0,6,0), (0,7,0) \right\},\\[0.2ex]
R_{\diamondsuit}(6,0,7)=R_{\diamondsuit}(6,0,4)\\[0.2ex]
\phantom{R_{\diamondsuit}(6,0,7)} = \left\{(0,0,0), (6,0,7), (6,0,4), (6,0,0), (0,0,4), (6,0,6), (0,0,6), (0,0,7) \right\},\\[0.2ex]
R_{\diamondsuit}(7,0,6)=R_{\diamondsuit}(4,0,6)\\[0.2ex]
\phantom{R_{\diamondsuit}(7,0,6)} = \left\{(0,0,0), (7,0,6), (4,0,6), (0,0,6), (4,0,0), (6,0,6), (6,0,0), (7,0,0) \right\},\\[0.2ex]
R_{\diamondsuit}(6,7,0)=R_{\diamondsuit}(6,4,0)\\[0.2ex]
\phantom{R_{\diamondsuit}(6,7,0)} = \left\{(0,0,0), (6,7,0), (6,4,0), (6,0,0), (0,4,0), (6,6,0), (0,6,0), (0,7,0) \right\},\\[0.2ex]
R_{\diamondsuit}(7,6,0)=R_{\diamondsuit}(4,6,0)\\[0.2ex]
\phantom{R_{\diamondsuit}(7,6,0)} = \left\{(0,0,0), (7,6,0), (4,6,0), (0,6,0), (4,0,0), (6,6,0), (6,0,0), (7,0,0) \right\},\\[3ex]
R_{\diamondsuit}(4,6,7)=R_{\diamondsuit}(7,6,4)\\[0.2ex]
\phantom{R_{\diamondsuit}(4,6,7)} = \left\{(0,0,0), (4,6,7), (7,6,4), (6,6,0), (4,0,4), (0,6,6), (6,0,6), (7,0,7) \right\},\\[0.2ex]
R_{\diamondsuit}(4,7,6)=R_{\diamondsuit}(7,4,6)\\
\phantom{R_{\diamondsuit}(4,7,6)} = \left\{(0,0,0), (4,7,6), (7,4,6), (6,0,6), (4,4,0), (0,6,6), (6,6,0), (7,7,0) \right\},\\
R_{\diamondsuit}(6,4,7)=R_{\diamondsuit}(6,7,4)\\[0.2ex]
\phantom{R_{\diamondsuit}(6,4,7)} = \left\{(0,0,0), (6,4,7), (6,7,4), (6,6,0), (0,4,4), (6,0,6), (0,6,6), (0,7,7) \right\},\\[0.2ex]
R_{\diamondsuit}(4,4,6)=R_{\diamondsuit}(7,7,6)\\[0.2ex]
\phantom{R_{\diamondsuit}(4,4,6)} = \left\{(0,0,0), (4,4,6), (7,7,6), (6,6,6), (4,4,0), (0,0,6), (6,6,0), (7,7,0) \right\},\\[0.2ex]
R_{\diamondsuit}(4,6,4)=R_{\diamondsuit}(7,6,7)\\[0.2ex]
\phantom{R_{\diamondsuit}(4,6,4)} = \left\{(0,0,0), (4,6,4), (7,6,7), (6,6,6), (4,0,4), (0,6,0), (6,0,6), (7,0,7) \right\},\\[0.2ex]
R_{\diamondsuit}(6,4,4)=R_{\diamondsuit}(6,7,7)\\[0.2ex]
\phantom{R_{\diamondsuit}(6,4,4)} = \left\{(0,0,0), (6,4,4), (6,7,7), (6,6,6), (0,4,4), (6,0,0), (0,6,6), (0,7,7) \right\},\\
R_{\diamondsuit}(0,4,7)=R_{\diamondsuit}(0,7,4)\\
\phantom{R_{\diamondsuit}(0,4,7)} = \left\{(0,0,0), (0,4,7), (0,7,4), (0,6,0), (0,4,4), (0,0,6), (0,6,6), (0,7,7) \right\},\\[0.2ex]
R_{\diamondsuit}(4,0,7)=R_{\diamondsuit}(7,0,4)\\[0.2ex]
\phantom{R_{\diamondsuit}(4,0,7)} = \left\{(0,0,0), (4,0,7), (7,0,4), (6,0,0), (4,0,4), (0,0,6), (6,0,6), (7,0,7) \right\},\\[0.2ex]
R_{\diamondsuit}(4,7,0)=R_{\diamondsuit}(7,4,0)\\[0.2ex]
\phantom{R_{\diamondsuit}(4,7,0)} = \left\{(0,0,0), (4,7,0), (7,4,0), (6,0,0), (4,4,0), (0,6,0), (6,6,0), (7,7,0) \right\},\\[2ex]
R_{\diamondsuit}(4,4,7)=R_{\diamondsuit}(7,7,4)\\[0.2ex]
\phantom{R_{\diamondsuit}(4,4,7)} = \left\{(0,0,0), (4,4,7), (7,7,4), (6,6,0), (4,4,4), (0,0,6), (6,6,6), (7,7,7) \right\},\\[0.2ex]
R_{\diamondsuit}(4,7,4)=R_{\diamondsuit}(7,4,7)\\[0.2ex]
\phantom{R_{\diamondsuit}(4,7,4)} = \left\{(0,0,0), (4,7,4), (7,4,7), (6,0,6), (4,4,4), (0,6,0), (6,6,6), (7,7,7) \right\},\\[0.2ex]
R_{\diamondsuit}(7,4,4)=R_{\diamondsuit}(4,7,7)\\[0.2ex]
\phantom{R_{\diamondsuit}(7,4,4)} = \left\{(0,0,0), (7,4,4), (4,7,7), (0,6,6), (4,4,4), (6,0,0), (6,6,6), (7,7,7) \right\},
\end{gather*}

\begin{figure}[t]
\centerline{\includegraphics[width=13.7cm]{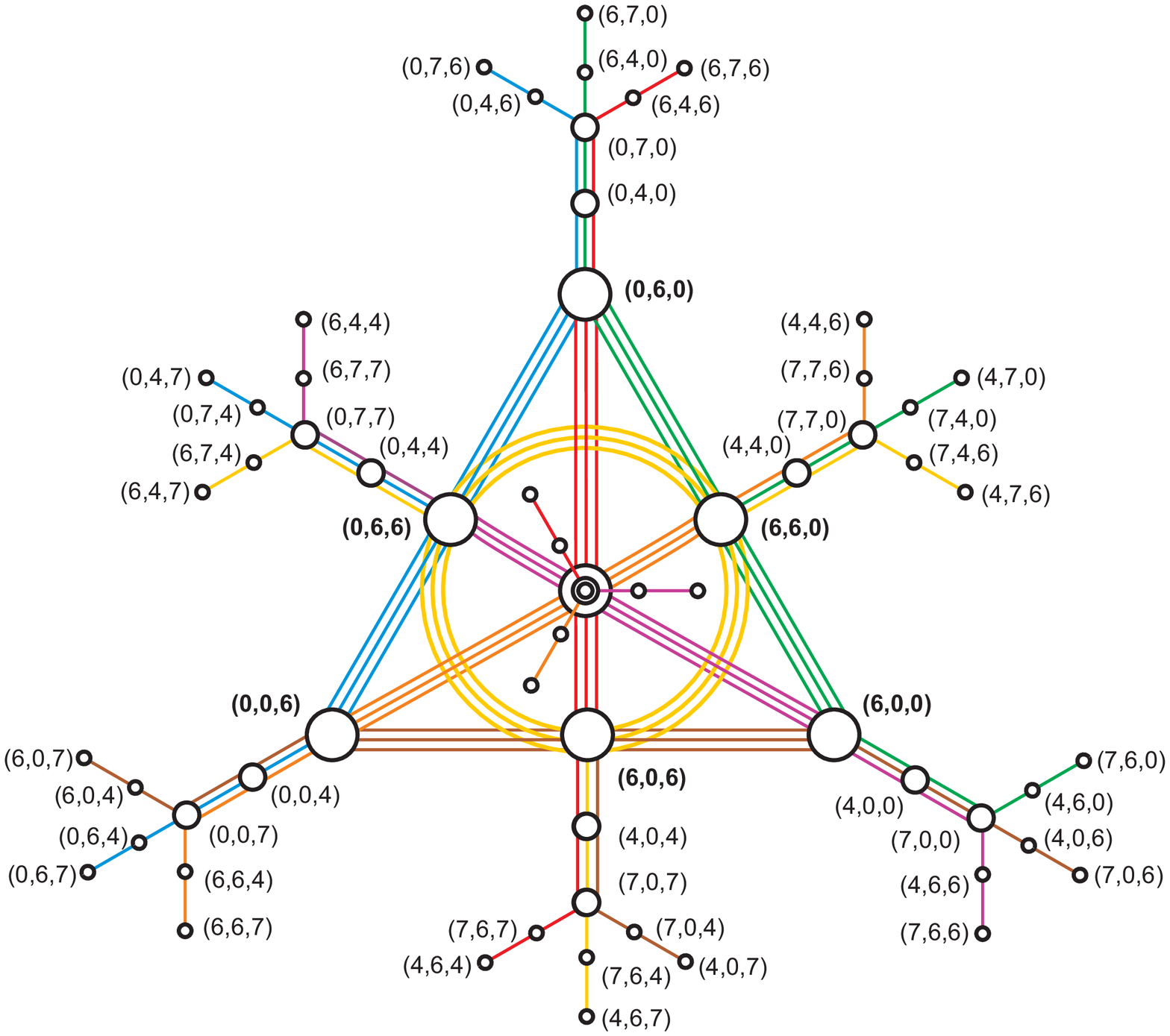}}

\caption{The Fano-Snowf\/lake -- a diagrammatic
illustration of a very intricate relation between the 21 free left
cyclic submodules generated by non-unimodular vectors of
$R_{\diamondsuit}^{3}$. Each circle represents a vector of
$R_{\diamondsuit}^{3}$ (in fact, of $I_{1}^{3}$), its size being roughly proportional to the
number of submodules passing through the given vector. As the
$(0,0,0)$ triple is not shown, each submodule is represented by
seven circles (three big, two medium-sized and two small) lying on
a common polygonal path. The small circles stand for the vectors generating the submodules.
The big circles represent the vectors with all
three entries from $J$; these vectors correspond to
the points of the Fano plane. In the middle ``branch'' there are two medium-size circles in front of the big one; in order to avoid a too crowded appearance of the f\/igure,
these and the associated six small circles are not given the corresponding vector labels.   The seven colors were chosen in such a way
to also make the lines of the Fano plane, i.e., the intersections of the submodules with~$J^3$, readily discernible. See~\cite{shpp} and/or~\cite{hs} for more details.}\label{fig1}
\end{figure}

\begin{figure}[t]
\centerline{\includegraphics[width=8.3truecm]{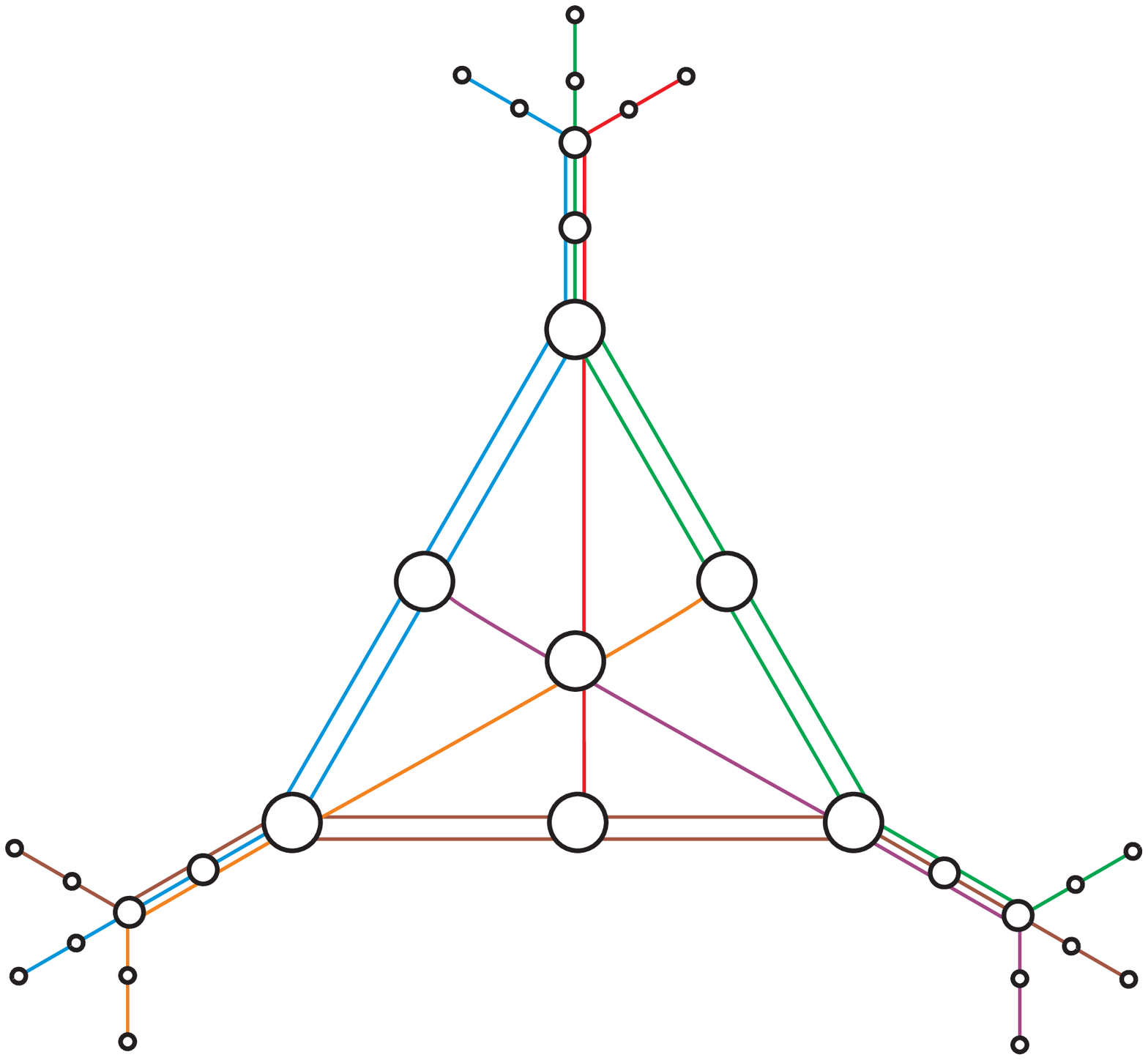}}
\vspace*{.1cm}
\centerline{\includegraphics[width=6.2truecm]{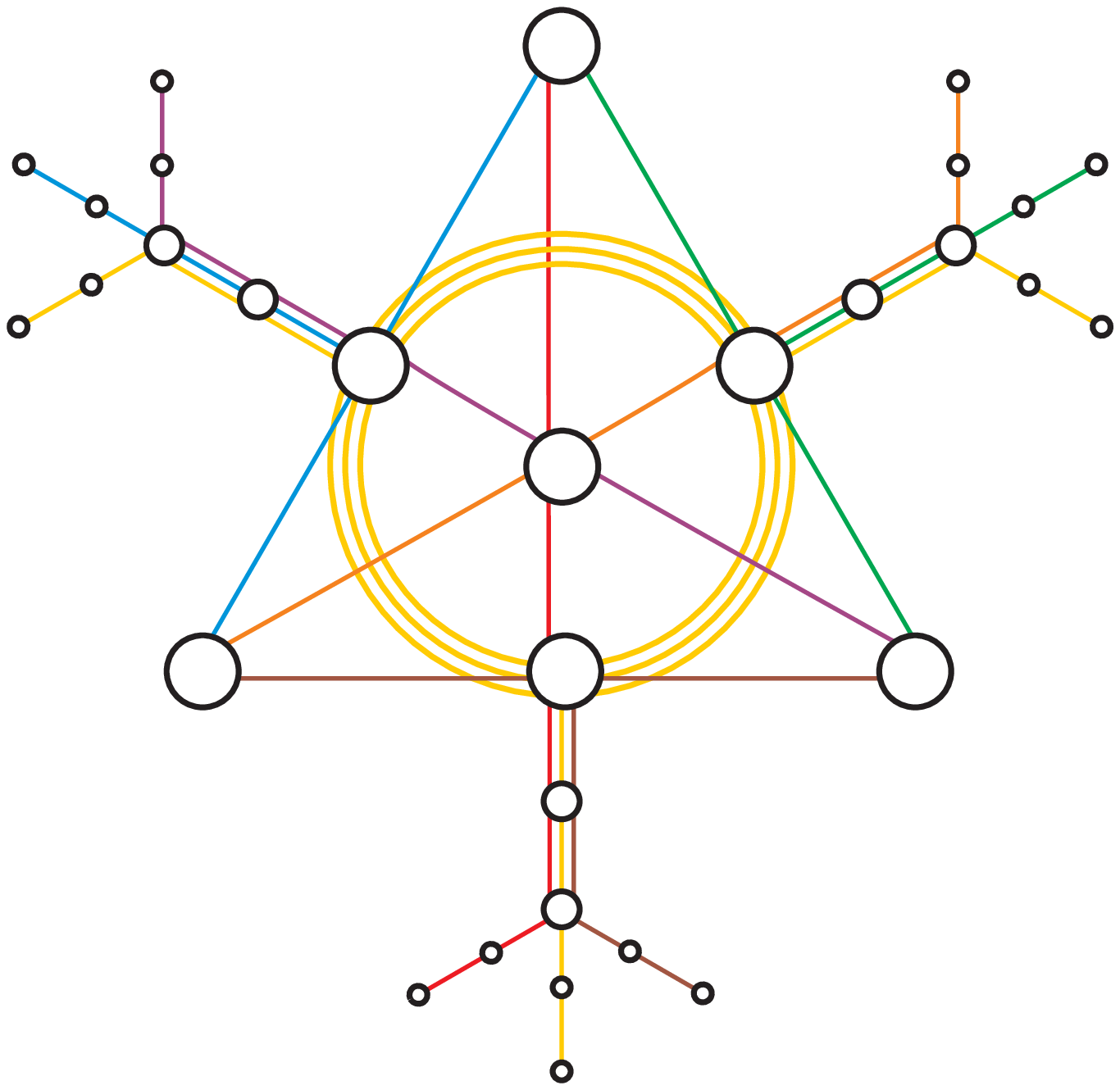}}
\vspace*{.1cm}
\centerline{\includegraphics[width=4.4truecm]{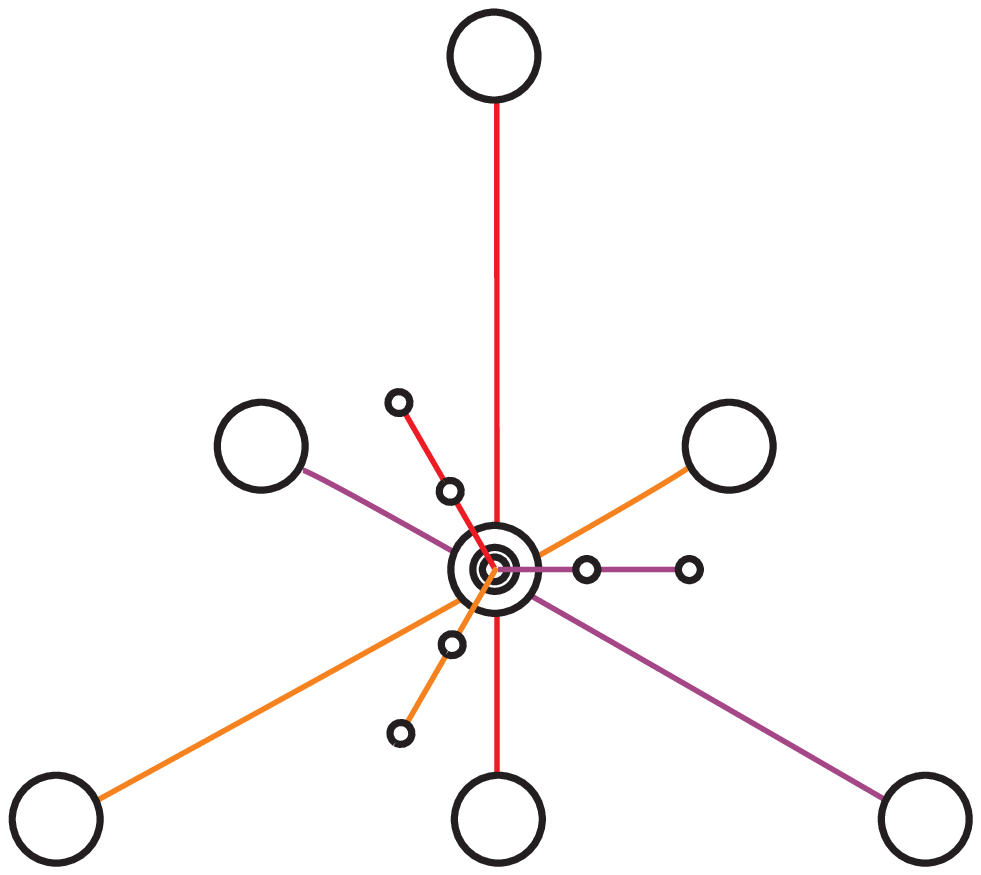}}

\caption{An illustration of the 9 -- 9 -- 3 decomposition of the set of  free cyclic submodules comprising the Fano-Snowf\/lake with respect to the Jacobson radical of $R_{\diamondsuit}$ according as the number
of radical entries in the submodule's generating vector(s) is two (top), one (middle) or zero (bottom), respectively.}\label{fig2}
\end{figure}
\begin{figure}[t]
\centerline{\includegraphics[width=13truecm]{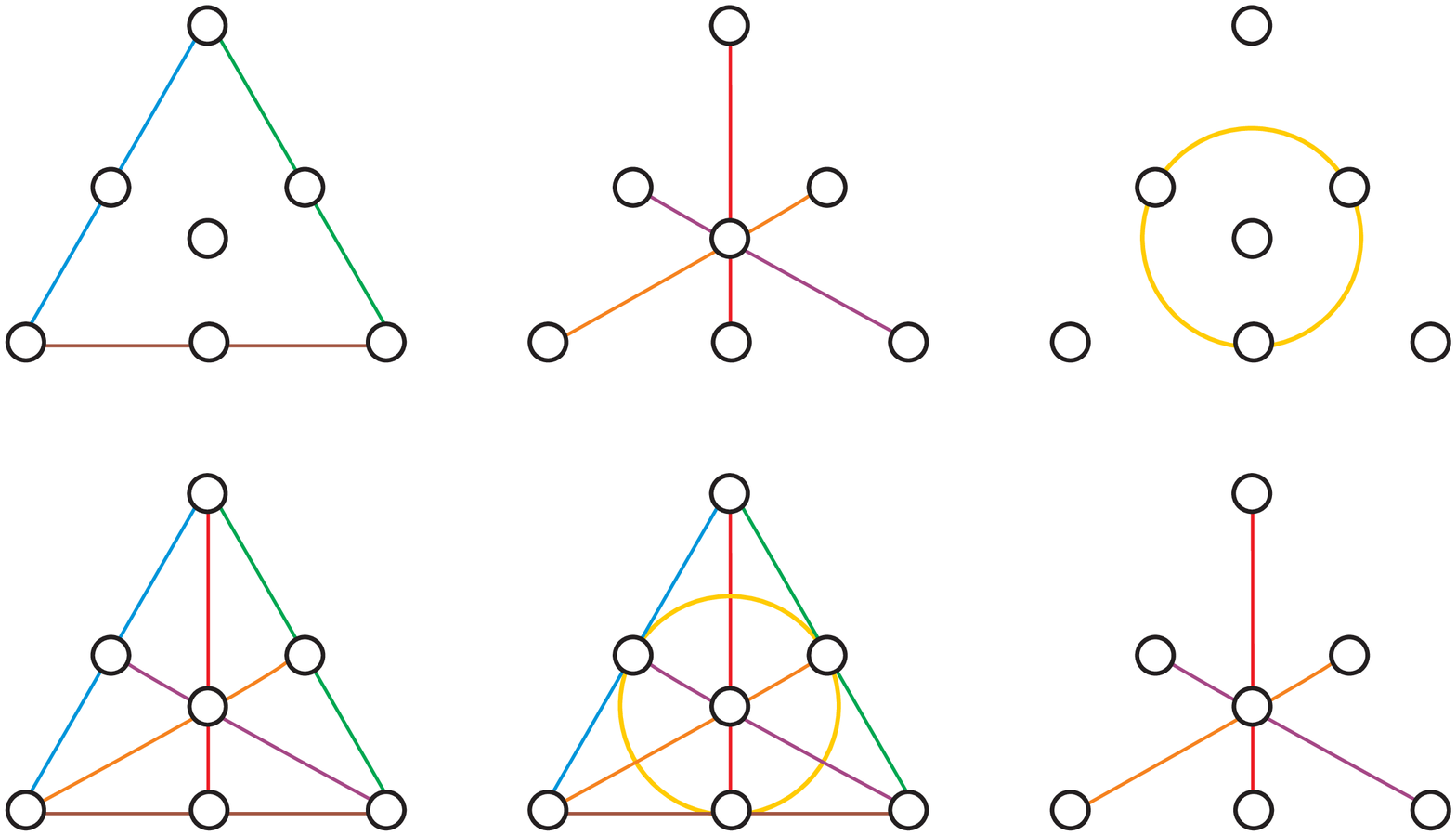}}

\caption{A comparison of the ``ternion-induced'' 6 -- 7 -- 3 factorization of the lines of the Fano plane (bottom) with the ordinary 3 -- 3 -- 1 one (top). Note a principal
qualitative dif\/ference between the two factorizations, since the three sets (factors) are pairwise disjoint in the latter case but not in the former one.}\label{fig3}
\end{figure}


\noindent
according as the number of Jacobson
radical entries in the generating vector(s) is two, one or zero,
respectively. Employing the picture of the Fano-Snowf\/lake given in
\cite{shpp} (reproduced, for convenience, in Fig.~\ref{fig1}), the
structure of and relation between the three sets can be
represented diagrammatically as shown in Fig.~\ref{fig2}. As each
submodule corresponds to a single line of the associated core Fano
plane, the decomposition of the Fano-Snowf\/lake induces an
intriguing factorization of the lines of the plane itself. This is
depicted in Fig.~\ref{fig3}, bottom panel, and it is seen to
fundamentally dif\/fer from the corresponding partitioning of the
Fano plane with respect to the Jacobson radical of its ground
f\/ield $GF(2)(\{0\}$), namely
\begin{gather*}
{GF}(2)(1,0,0) = \left\{ (0,0,0), (1,0,0) \right\},\qquad
{GF}(2)(0,1,0) = \left\{ (0,0,0), (0,1,0) \right\},\\
{GF}(2)(0,0,1) = \left\{ (0,0,0), (0,0,1) \right\},\qquad
{GF}(2)(1,1,0) = \left\{ (0,0,0), (1,1,0) \right\},\\
{GF}(2)(1,0,1) = \left\{ (0,0,0), (1,0,1) \right\},\qquad
{GF}(2)(0,1,1) = \left\{ (0,0,0), (0,1,1) \right\},\\
{GF}(2)(1,1,1) = \left\{ (0,0,0), (1,1,1) \right\},
\end{gather*}
as displayed in Fig.~\ref{fig3}, top panel.

The origin of the factorization of the Fano plane when related to
its ground f\/ield $GF(2)$ is easy to understand: the number of the
Jacobson radical entries (i.e., only zeros in this case) in the
coordinates of a line (and, by duality, of a point as well) has a
clear meaning with respect to the triangle of base points of the
coordinate system. Something similar holds obviously for the
factorization of the Fano-Snowf\/lake with respect to its ternionic
coordinates, but passing to the embedded Fano plane this link
seems to be lost or substantially distorted. Fig.~\ref{fig2} illustrates
this fact quite nicely: in the top f\/igure the maximum number of polygonal paths pass
through the three corners/vertices of the basic triangle, in the
middle f\/igure this property is enjoyed by the points on each side of
the triangle which are not vertices, whereas in the bottom f\/igure
all the branches share the single point which is out of the reference
triangle. It is the intersections of the branches/polygonal paths with
the core Fano plane which behave ``strangely'' and give rise to the
fundamental dif\/ference between the two factorizations of the Fano
plane shown in Fig~\ref{fig3}.

The observations above clearly demonstrate that there is
more to the algebraic structure of the Fano plane than meets the
eye. The plane when considered on {\it its own} is found to
``reveal'' quite dif\/ferent aspects compared with the case when
{\it embedded into} a more general, non-unimodular
projective lattice setting. This dif\/ference is likely to get more
pronounced, and more intricate as well, as we pass to higher order
rings giving rise to more complex forms of Fano-Snowf\/lakes. A key
question is to f\/ind out whether the Snowf\/lakes' decomposition
patterns and their induced factorizations of the lines of the core
Fano planes remain qualitatively the same as in the ternionic
case; our preliminary analysis of such structures over a
particular class of non-commutative rings of order sixteen and
having twelve zero-divisors indicates that this might be so.
Another line of exploration worth pursuing is to stay with
ternions but focus on higher-order ($ q > 2$) and/or
higher-dimensional ($n > 2$) ``Snowf\/lake'' geometries and their
core projective planes and/or spaces. Finding, however, a general
construction principle for these remarkable geometrical structures
with respect to the properties of def\/ining rings currently seems to be~-- already for the simplest $q=2$ case~-- a truly dif\/f\/icult, yet extremely challenging task due to the ``ubiquity''
of the Fano plane in various mathematical and physical contexts (see, e.g.,~\cite{br}).

\subsection*{Acknowledgements}

The work was partially supported by the VEGA grant agency projects
Nos. 6070 and 7012, the CNRS-SAV Project No.~20246 and
by the Action Austria--Slovakia project No. 58s2. We thank Hans Havlicek
(Vienna University of Technology) for valuable comments and suggestions.

\pdfbookmark[1]{References}{ref}
\LastPageEnding

\end{document}